\documentclass{book}    
\usepackage{piers}  
\usepackage{graphicx}
\begin{document}

\title{Analysis of multipactor effect in parallel-plate 
and rectangular waveguides}
\maketitle

\author      {A. Berenguer}
\affiliation {Departamento de Ingenier\'ia de Comunicaciones, Universidad Miguel Hern\'andez de Elche}
\address     {Avenida de la Universidad s/n}
\city        {Elche}
\postalcode  {03202}
\country     {Spain}
\phone       {}    
\fax         {}    
\email       {aberenguer@umh.es}  
\misc        { }  
\nomakeauthor

\author      {A. Coves}
\affiliation {Departamento de Ingenier\'ia de Comunicaciones, Universidad Miguel Hern\'andez de Elche}
\address     {Avenida de la Universidad s/n}
\city        {Elche}
\postalcode  {03202}
\country     {Spain}
\phone       {+34966658415}    
\fax         {}    
\email       {angela.coves@umh.es}  
\misc        { }  
\nomakeauthor
\author      {E. Bronchalo}
\affiliation {Departamento de Ingenier\'ia de Comunicaciones, Universidad Miguel Hern\'andez de Elche}
\address     {Avenida de la Universidad s/n}
\city        {Elche}
\postalcode  {03202}
\country     {Spain}
\phone       {+34966658593}    
\fax         {}    
\email       {ebronchalo@umh.es}  
\misc        { }  
\nomakeauthor
\author      {B. Gimeno}
\affiliation {Departamento de F\'isica Aplicada y Electromagnetismo - Instituto de Ciencia de Materiales. Universidad de Valencia}
\address     {Av. de Blasco Ib\'añez, 13}
\city        {Valencia}
\postalcode  {46010}
\country     {Spain}
\phone       {}    
\fax         {}    
\email       {benito.gimeno@uv.es}  
\misc        { }  
\nomakeauthor
\author      {V. Boria}
\affiliation {Departamento de Comunicaciones, Universidad Polit\'ecnica de Valencia}
\address     {Camino de Vera s/n}
\city        {Valencia}
\postalcode  {46022}
\country     {Spain}
\phone       {}    
\fax         {}    
\email       {vboria@dcom.upv.es}  
\misc        { }  
\nomakeauthor

\begin{authors}

{\bf A. Berenguer}$^{1}$, {\bf A. Coves}$^{1}$, {\bf E. Bronchalo}$^{1}$, {\bf B. Gimeno}$^{2}$, {\bf and V. Boria}$^{3}$\\
\medskip
$^{1}$Departamento de Ingenier\'ia de Comunicaciones. Universidad Miguel Hern\'andez de Elche, Spain\\
$^{2}$Departamento de F\'isica Aplicada y Electromagnetismo - Instituto de Ciencia de Materiales. Universidad de Valencia, Spain\\
$^{3}$Departamento de Comunicaciones, Universidad Polit\'ecnica de Valencia, Spain

\end{authors}

\begin{paper}

\begin{piersabstract}
In this work it is investigated, in the parallel-plate waveguide case, how the 1D, 2D or 3D motion of the electrons inside the waveguide can affect the generalized susceptibility diagrams, by means of a developed model capable of tracking the exact trajectory of multiple effective electrons which includes effects such as the spreading of the secondary electron departure kinetic energies or the dependence on elastic and inelastic electrons. On the other hand, a comparative study of the susceptibility charts in a parallel-plate and in its equivalent rectangular waveguide with the same height is performed, showing how the inhomogeneity of the electric field inside the waveguide modifies the multipactor region with respect to that predicted by the parallel-plate waveguide case. 
The results of this study are going to be extended to a partially dielectric-loaded rectangular waveguide, which is a problem of great interest in the space industry that has not yet been rigorously investigated in the literature. 
\end{piersabstract}

\psection{Introduction}
The multipactor effect is an electron discharge that may appear in particle accelerators and microwave devices such as waveguides in satellite on-board equipment under vacuum conditions. This effect has been widely studied, and many investigations have been focused on the prediction of multipactor breakdown in a wide variety of microwave passive components for signals of different frequency and power levels, in order to prevent their damage. Many works \cite{Esteban, Quesada} take advantage of available susceptibility charts in empty parallel-plate waveguides obtained with analytical models \cite{Hatch}, and they are directly used to predict multipactor breakdown in the component under study, which is going to happen in the point of highest field intensity. Thus, they are aimed to determine such highest field intensity region, which is generally the smallest device gap. However, such susceptibility diagrams do not take into account important effects such as the dependence of these diagrams on elastic and inelastic electrons, as well as the 3D character of the motion of the electrons inside the waveguide, or the non-uniform nature of the electromagnetic fields in some particular cases.
In the last years, the authors have developed a 1D model for studying the multipactor effect in a parallel-plate dielectric-loaded waveguide \cite{Torregrosa06, Coves, Torregrosa10} including space charge effects. In this work we have extended this model to a 3D movement case, and a comparison of the obtained results with such 3D model to those provided by the 1D model has been performed in the empty parallel-plate waveguide case. On the other hand, a 3D model for studying multipactor effect in an empty rectangular waveguide has also been developed, and the obtained results for a rectangular waveguide with the same height as the parallel-plate one have also been obtained and compared.

\psection{Theory}
\psubsection{Electron Dynamics}
The key for understanding the mechanism of the multipactor discharge is to study the behavior of the electrons within a harmonically excited waveguide. An electron inside the gap space is accelerated by the RF electric field. A significant growth in the electron density in the device can only happen if the electrons hit the walls with the appropriate energy and at periodically suitable instants.

The electron dynamics is governed by the \textit{Lorentz force} resulting in Eq. (\ref {Lorentz}),
\begin{equation} \label{Lorentz}
	\vec{F_L} = q (\vec{E}+\vec{v} \times \vec{B}) = \frac{\partial \vec{p}}{\partial t}
\end{equation}
where $q=-e$ is the electron charge, $\vec{E}$ and $\vec{B}$ are, respectively, the instantaneous electric
and magnetic field vectors interacting with the electron, $\vec{v}$ is the velocity vector of the electron, and $t$ is the time. The linear momentum is defined according to Eq. (\ref{momentum}),
\begin{equation} \label{momentum}
	\vec{p} = m_0\gamma\vec{v}
\end{equation}
where $m_0$ is the electron mass at rest, $\gamma = \frac{1}{\sqrt{1-\left({\frac{v}{c}}\right)^2}}$ is the relativistic factor, $v$ being the magnitude of the velocity vector, $c = \frac{1}{\sqrt{\mu_0\epsilon_0}}$ is the free-space light velocity, where $\mu_0$ is the free-space magnetic permeability and $\epsilon_0$ is the free-space electric permittivity. Usually, when $v \ll c$, the relativistic term tends to  $\gamma \rightarrow 1$. Although the relativistic component of this equation can be discarded for the typical power ranges of most space waveguide devices, it should be considered in cases when extreme speeds are reached $\left(\frac{v}{c}\gtrsim 0.1\right)$, like for high-power multipactor testing simulations. Expanding then Eq. (\ref{Lorentz}), Eq. (\ref{Lorentz_momentum}) is obtained:
\begin{equation} \label{Lorentz_momentum}
	-\vec{E}-\vec{v} \times \vec{B} = A \gamma \vec{a} + \frac{A}{c^2}\gamma^3(\vec{v}\cdot\vec{a})\vec{v}
\end{equation}
where $\vec{a}$ is the acceleration vector and $A = m_0/e$. The acceleration is the derivative over
time of the velocity, which in turn is the derivative of the position: $\vec{a} = \dot{\vec{v}} = \ddot{\vec{r}}$. The differential equation system to solve becomes then,

\begin{subequations}\label{}
\begin{equation} \label{}
\ddot{r}_x = \frac{\dot{r}_z B_y-\dot{r}_y B_z-E_x+\dot{r}_x \cdot\dot{\vec{r}}\cdot\frac{\vec{E}}{c^2}}{A\gamma}
\end{equation}
\begin{equation} \label{}
\ddot{r}_y = \frac{\dot{r}_x B_z-\dot{r}_z B_x-E_y+\dot{r}_y \cdot\dot{\vec{r}}\cdot\frac{\vec{E}}{c^2}}{A\gamma}
\end{equation}
\begin{equation} \label{}
\ddot{r}_z = \frac{\dot{r}_y B_x-\dot{r}_x B_y-E_z+\dot{r}_z \cdot\dot{\vec{r}}\cdot\frac{\vec{E}}{c^2}}{A\gamma}
\end{equation}
\end{subequations}

In the parallel-plate dielectric-loaded waveguide case, the electromagnetic fields have an analytical expression, and then such differential equation system can be analytically solved under certain approximations \cite{Torregrosa06, Coves}. On the other side, in a partially dielectric-loaded rectangular waveguide, the electromagnetic fields must be numerically solved. To this end, a vectorial modal method has been employed \cite{coves_tesis}, and in this case the electron trajectories are found by solving such equations using a velocity Verlet algorithm which assures sufficient accuracy and good efficiency provided enough time steps are chosen.

\psubsection{Secondary Electron Yield (SEY)}
The movement may eventually lead the electron to impact with any surface. Each collision can result in the emission or absorption of secondary electrons. The number of electrons emitted or absorbed after each impact is determined by the value of the secondary electron yield (SEY) parameter $\delta$ ($\delta > 1$ if secondary electrons are emitted, and $\delta < 1$ if they are absorbed). The value of parameter $\delta$ is calculated by means of realistic SEY functions that include the effect of reflected electrons for low impact energies of primary electrons, which must be accounted in order to obtain accurate results \cite{Vicente,Seviour}. The SEY function used assumes $\delta = 0.5$ for low impact energies of primary electrons. This value is in agreement with experimental results obtained in \cite{Cimino}. If total absorption were considered instead, an electron colliding with low energy would automatically be lost for the rest of the simulation. The SEY properties are defined by the following parameters: the primary electron impact kinetic energies, which yield $\delta = 1$, $W_1$, and $W_2$; the impact energy $W_{max}$ necessary for a primary electron to yield $\delta = \delta_{max}$, which is the maximum value of the SEY function; and the value of the primary electron impact energy $W_0 (\delta = 0)$ that limits the region of elastic collisions. In the effective electron model assumed in this study, after each effective electron impacts at time $t$ with any surface, $N_i(t)$ is modified according to the $\delta$ value provided by the SEY function by means of Eq. (\ref{Ni(t)}),
\begin{equation} \label{Ni(t)}
	N_i(t+\Delta t) = \delta N_i(t)
\end{equation}
where $N_i(t)$ represents the population of each effective electron inside the waveguide at the instant $t$, and $\Delta t$ is the time step used in the simulations. If multiple effective electrons are considered, the total population of electrons inside the waveguide at each instant $t$ is the sum of the population of all effective electrons. 

On the other hand, the secondary electron departure kinetic energy $W_s$ after each impact is assumed to follow a Rayleigh probability density function according to Eq. (\ref{f(W_s)}) \cite{Torregrosa06},
\begin{equation} \label{f(W_s)}
	f(W_s) = \frac{W_s}{W_g^2}e^{\frac{-W_s^2}{2W_g^2}}
\end{equation}
with parameter $W_g = 3$ eV.

\psection{Numerical results and discussion}
The schemes of the geometries and dimensions of the problems under investigation are shown in Fig. \ref{geometry}.
\begin{figure}[h]
\centering
\subfigure[Parallel-plate waveguide.]{\includegraphics[width=0.45\textwidth]{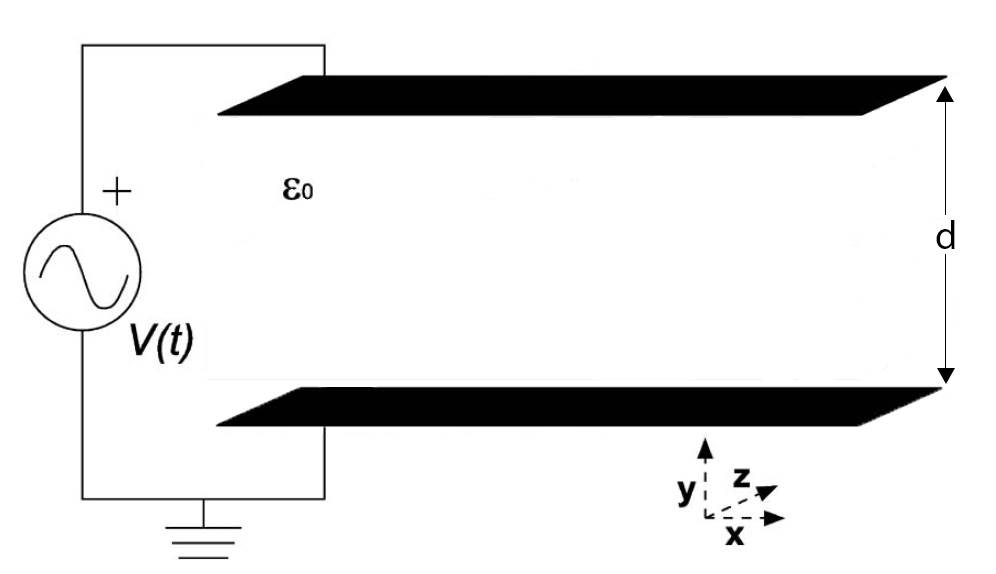}}
\subfigure[Rectangular waveguide.]{\includegraphics[width=0.45\textwidth]{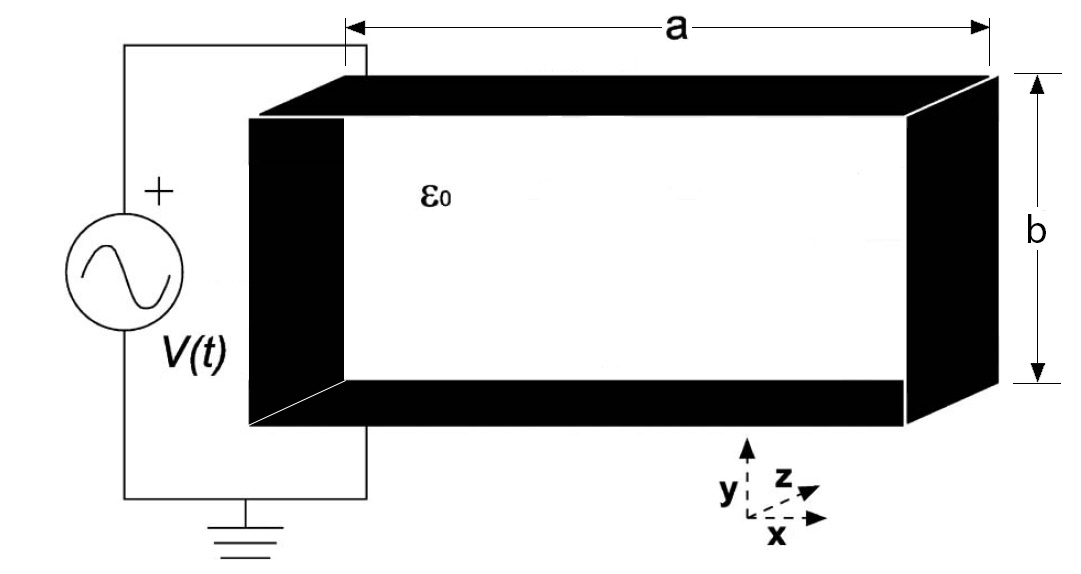}}
\caption{Geometries and dimensions of the problems under investigation.} \label{geometry}
\end{figure}
For the parallel-plate waveguide the distance is $d=3mm$. Similarly, in the case of the rectangular waveguide, the distance is $b=3mm$. With regard to the distance $a$, it has been chosen so that there is at least one propagation mode in the waveguide, i.e.  $a \gg b$. Fig. \ref{SEY_silver} provides the SEY characteristics for the silver material employed in the waveguides under study. The parameters are $W_1 = 30$ eV, $W_2 = 5000$ eV, $W_{max} = 165$ eV, $W_0 = 16$ eV, and $\delta_{max} = 2.22$.

\begin{figure}[h]
	\centering
		\includegraphics[width=0.4\textwidth]{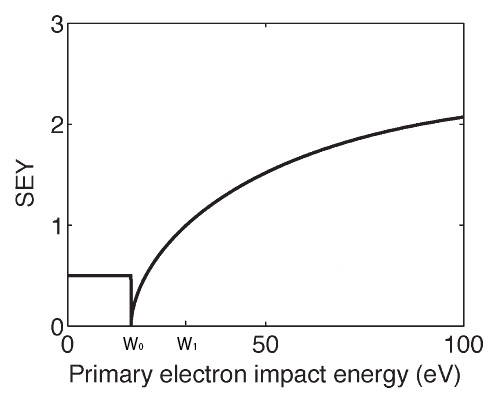}
	\caption{SEY characteristics for the silver material.}\label{SEY_silver}
\end{figure}

First, a comparative study of the susceptibility charts in a parallel-plate and in its equivalent rectangular waveguide with the same height is performed. In each susceptibility chart, for each $V_0$ and $f \times d$ pair, the simulation is run 72 times, corresponding to 72 equidistant phases of the RF field separated 5 degrees. In each run, a single effective electron starts at a uniformly random position and velocity inside the waveguide. Each simulation lasts 100 RF cycles, with 5000 time intervals per cycle. The arithmetic mean of the final population of electrons after 100 RF cycles is calculated using all 72 simulations. If this mean value is greater than 1, then the multipactor discharge is assumed to have occurred. In Fig. \ref{cartas susceptibilidad}, the susceptibility charts of an empty silver parallel-plate waveguide considering both a 1D model and a 3D model have been compared (results obtained with a 2D model have also been obtained for this guide, although they are not shown in this study given that the results of 2D and 3D models are quite similar).
\begin{figure}[h]
\centering
\subfigure[1D: parallel-plate waveguide.]{\includegraphics[width=0.3\textwidth]{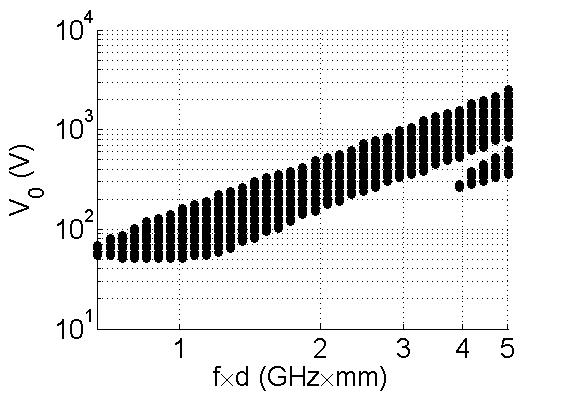}}
\subfigure[3D: parallel-plate waveguide.]{\includegraphics[width=0.3\textwidth]{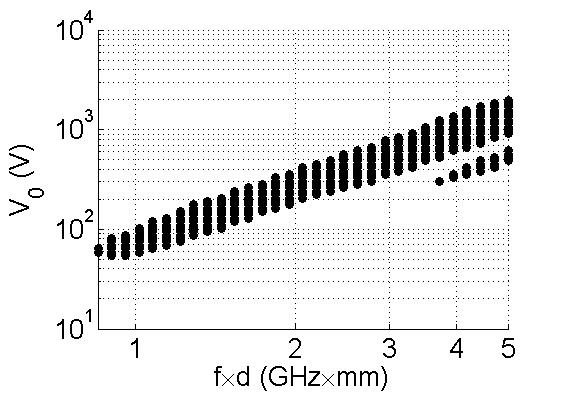}}
\subfigure[3D: rectangular waveguide.]{\includegraphics[width=0.3\textwidth]{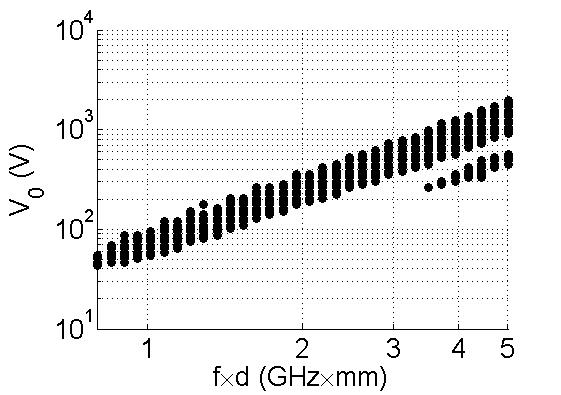}}
\caption{Susceptibility charts.} \label{cartas susceptibilidad}
\end{figure}
The differences between the susceptibility charts shown in Fig. \ref{cartas susceptibilidad} are evident: the multipactor region experiences a relevant decrease when the 3D movement inside the waveguide is considered. Such decrease is observed not only as a multipactor threshold reduction, i.e., a decrease of the voltage threshold at each $f \times d$ point, but also in the high voltage regions. This observed decrease of the multipactor region in a 3D movement case is due to a slight desynchronization of the electron resonant movement after each impact, given that the secondary electron departure kinetic energy $W_s$ is distributed in this case in the three velocity components (in a 1D model, all the secondary electron departure kinetic energy is used to push the effective electron towards the opposite wall). In the case of the 3D rectangular waveguide model, the susceptibility chart seems very similar to the 3D parallel-plate waveguide, i.e. a relevant decrease in the multipactor region can be appreciated when compared to the 1D parallel-plate waveguide model.

Secondly, in order to perform a detailed analysis of the physical behavior and the dynamics of the electron inside the waveguides under study, different parameters have been compared considering a bias point within the multipactor region. In all cases, $V_0 = 70V$, $f \times d = 1 GHz \times mm$ is the point under analysis. 

In Fig. \ref{n_electrons} the populations of electrons after 100 RF cycles are shown.	For the 1D parallel-plate waveguide model, the population of electrons rises steadily until a saturation level around $10^{10}$ electrons, which is reached (due to space-charge effect) after 25 RF cycles approximately. Similar results are obtained in the 3D parallel-plate waveguide model. However, the saturation level is reached after 50 RF cycles in this case. With regard to the 3D rectangular waveguide model, no saturation level can be appreciated after 100 RF cycles since no space-charge effect has been considered in this guide. However, the same population level as in the 3D  parallel-plate waveguide model is achieved again after 45 RF cycles, which corroborates that in this 3D model there is a desynchronization of the electron resonant movement after each impact.
\begin{figure}[h]
\centering
\subfigure[1D: parallel-plate waveguide.]{\includegraphics[width=0.3\textwidth]{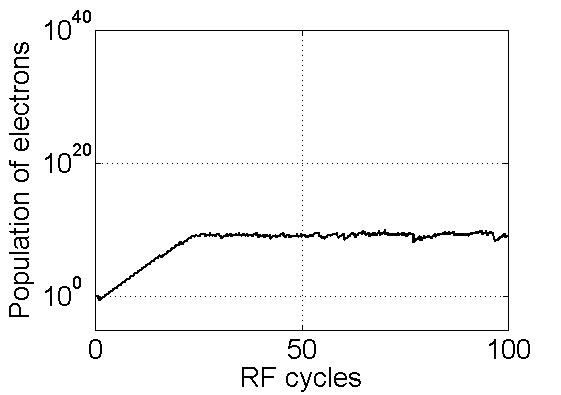}}\label{n_electrons_1D_ppp}
\subfigure[3D: parallel-plate waveguide.]{\includegraphics[width=0.3\textwidth]{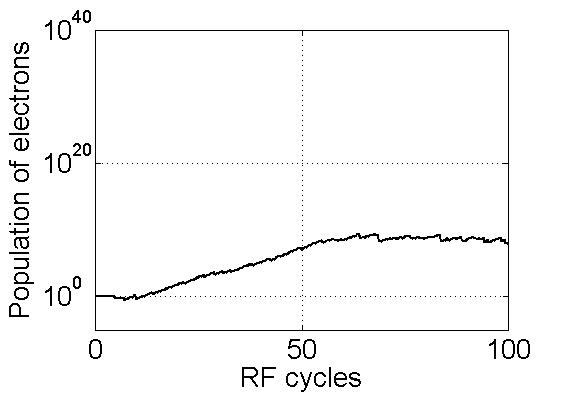}}\label{n_electrons_3D_ppp}
\subfigure[3D: rectangular waveguide.]{\includegraphics[width=0.3\textwidth]{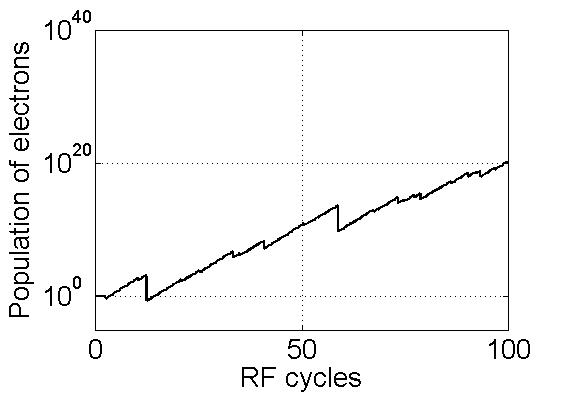}}\label{n_electrons_3D_guia}
\caption{Populations of electrons.} \label{n_electrons}
\end{figure}

Finally, the $y$ position of the electron after 30 RF cycles is shown in Fig. \ref{y position}. As it can be appreciated, both for the 3D parallel-plate and rectangular waveguide models, there is a slight desynchronization in the path followed by the electron, mainly on the first RF cycles.
\begin{figure}[h]
\centering
\subfigure[1D: parallel-plate waveguide.]{\includegraphics[width=0.3\textwidth]{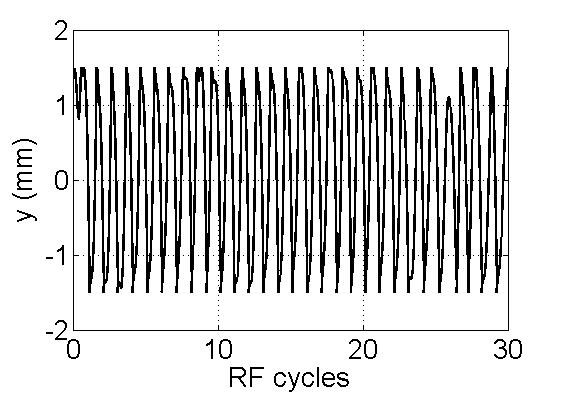}}
\subfigure[3D: parallel-plate waveguide.]{\includegraphics[width=0.3\textwidth]{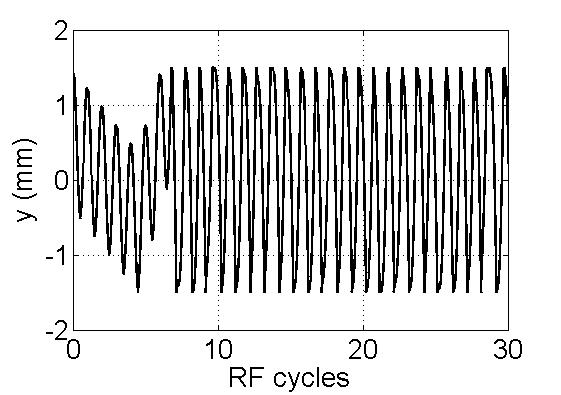}}
\subfigure[3D: rectangular waveguide.]{\includegraphics[width=0.3\textwidth]{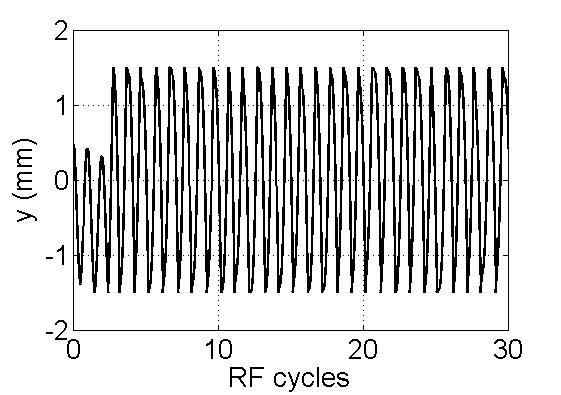}}
\caption{$y$ position of the electron.} \label{y position}
\end{figure}

The results of this study are going to be extended to a partially dielectric-loaded rectangular waveguide, which is a problem of great interest in the space industry that has not yet been rigorously investigated in the literature.

\psection{Conclusion}
The 1D parallel-plate waveguide model, which is commonly used in the multipactor effect analysis in a wide variety of microwave passive components,is not accurate enough in the highest field intensity region. There is a x-coordinate dependency in the electromagnetic fields inside the waveguide which is not considered in the 1D parallel-plate waveguide model. In this work, a 3D model which takes into account this dependency has been shown, both for parallel-plate and rectangular waveguides. The results obtained are accurate when compared to the 1D model.

\ack
This work was supported by the Ministerio de Econom\'ia y Competitividad, Spanish Government, under the coordinated projects TEC2013-47037-C5-4-R and TEC2013-47037-C5-1-R.

\end{paper}

\end{document}